\documentclass[12pt,a4paper,tikz]{article}

\usepackage{aasymbols}
\usepackage{amssymb}
\usepackage{natbib}
\usepackage[usenames, dvipsnames]{color}
\usepackage{graphicx}
\usepackage{ragged2e}
\usepackage{hyperref}
\usepackage{txfonts}
\usepackage{ccite}

\usepackage[utf8]{inputenc}
\usepackage[english]{babel}

\newcommand{\Line}[3]{\Ion{#1}{#2}~#3}

\newcommand{\Ion}[2]{#1{\,\small\textsc{#2}}}

\usepackage{geometry}
\usepackage{fancyhdr}
\begin{document}

\raggedright
\large
Astro~2020 Science White Paper \linebreak
\textbf{Evolved Planetary Systems around White Dwarfs} \linebreak
\normalsize

\noindent \textbf{Thematic Areas:} \hspace*{60pt} $\square$ \hspace*{-9.5pt}x Planetary Systems \hspace*{10pt} $\square$ Star and Planet Formation \hspace*{20pt}\linebreak
$\square$ Formation and Evolution of Compact Objects \hspace*{31pt} $\square$ Cosmology and Fundamental Physics \linebreak
  $\square$  Stars and Stellar Evolution \hspace*{1pt} $\square$ Resolved Stellar Populations and their Environments \hspace*{40pt} \linebreak
  $\square$    Galaxy Evolution   \hspace*{45pt} $\square$             Multi-Messenger Astronomy and Astrophysics \hspace*{65pt} \linebreak
  
\textbf{Principal Author:}

Name: Boris G\"ansicke
 \linebreak						
Institution:  University of Warwick
 \linebreak
Email: Boris.Gaensicke@warwick.ac.uk
 \linebreak
Phone:  +44\,2476\,574741
 \linebreak
 
\justify
\textbf{Co-authors:} 
Martin Barstow (University of Leicester),
Amy Bonsor (University of Cambridge),
John Debes (STScI),
Patrick Dufour (Universit\'e de Montr\'eal),
Tim Cunningham (University of Warwick),
Erik Dennihy (Gemini Observatory),
Nicola Gentile Fusillo (University of Warwick),
Jay Farihi (UCL),
Mark Hollands (University of Warwick),
Matthew Hoskin (University of Warwick),
Paula Izquierdo (IAC),
Jennifer Johnson (OSU),
Beth Klein (UCLA),
Detlev Koester (University of Kiel),
Juna Kollmeier (Carnegie Observatories),
Wladimir Lyra (California State University at Northridge),
Christopher Manser (University of Warwick),
Carl Melis (UCSD),
Pablo Rodr{\'i}guez-Gil (IAC/Universidad de La Laguna),
Matthias Schreiber (Universidad de Valpara{\'i}so),
Andrew Swan (UCL),
Odette Toloza (University of Warwick),
Pier-Emmanuel Tremblay (University of Warwick),
Dimitri Veras (University of Warwick),
David Wilson (University of Texas), 
Siyi Xu (Gemini Observatory),
Ben Zuckerman (UCLA),

\medskip\noindent
\textbf{Abstract  (optional):}\\
Practically all known planet hosts will evolve into white dwarfs, and large parts of their planetary systems will survive this transition~--~the same is true for the solar system beyond the orbit of Mars. Spectroscopy of white dwarfs accreting planetary debris provides the most accurate insight into the bulk composition of exo-planets. Ground-based spectroscopic surveys of $\simeq260,000$ white dwarfs detected with \textit{Gaia} will identify $>1000$ evolved planetary systems, and high-throughput high-resolution space-based ultraviolet spectroscopy is essential to measure in detail their abundances. So far, evidence for two planetesimals orbiting closely around white dwarfs has been obtained, and their study provides important constraints on the composition and internal structure of these bodies. Major photometric and spectroscopic efforts will be necessary to assemble a sample of such close-in planetesimals that is sufficiently large to establish their properties as a population, and to deduce the architectures of the outer planetary systems from where they originated. Mid-infrared spectroscopy of the dusty disks will provide detailed mineralogical information of the debris, which, in combination with the elemental abundances measured from the white dwarf spectroscopy, will enable detailed physical modelling of the chemical, thermodynamic, and physical history of the accreted material. Flexible multi-epoch infrared observations are essential to determine the physical nature, and origin of the variability observed in many of the dusty disks. Finally, the direct detection of the outer reservoirs feeding material to the white dwarfs will require sensitive mid- and far-infrared capabilities. 

\pagebreak

\justify

\smallskip\noindent
\textbf{\large Overview}\\
Practically all known planet hosts share one common fate: they will evolve into white dwarfs, the embers of main-sequence stars with initial masses $\la8\,M_\odot$. Many of the known planets will survive the post main-sequence evolution of their host stars~--~including the solar system Mars and beyond \citep{schroeder+connonsmith08-1}. The gravitational interactions of these planets can scatter asteroids, moons, and possibly some of the planets themselves deep into the gravitational potential of the white dwarf, where they are tidally disrupted and  eventually accreted \citep{jura03-1, veras+gaensicke15-1, payneetal17-1}. Observational evidence for planetary systems at white dwarfs is ample in the form of photospheric contamination by the accreted debris \citep{zuckermanetal10-1, koesteretal14-1}, and dusty \citep{farihietal09-1} and gaseous circumstellar disks \citep{gaensickeetal06-3,manseretal16-1}. \cite{vanderburgetal15-1} and \cite{manseretal19-1} established the first photometric and spectroscopic evidence for exo-planetesimals in close orbits around white dwarfs. For recent review papers on evolved planetary systems at white dwarfs, see \citet{farihi16-1} and \citet{veras16-1}.
This white paper describes an ambitious research programme into the architectures of evolved planetary systems and their use as probes of the bulk abundances of exo-planetesimals, and we identify the facilities required over the next decade to reach our scientific goals.

\begin{figure*}
\centerline{\includegraphics[width=0.9\textwidth]{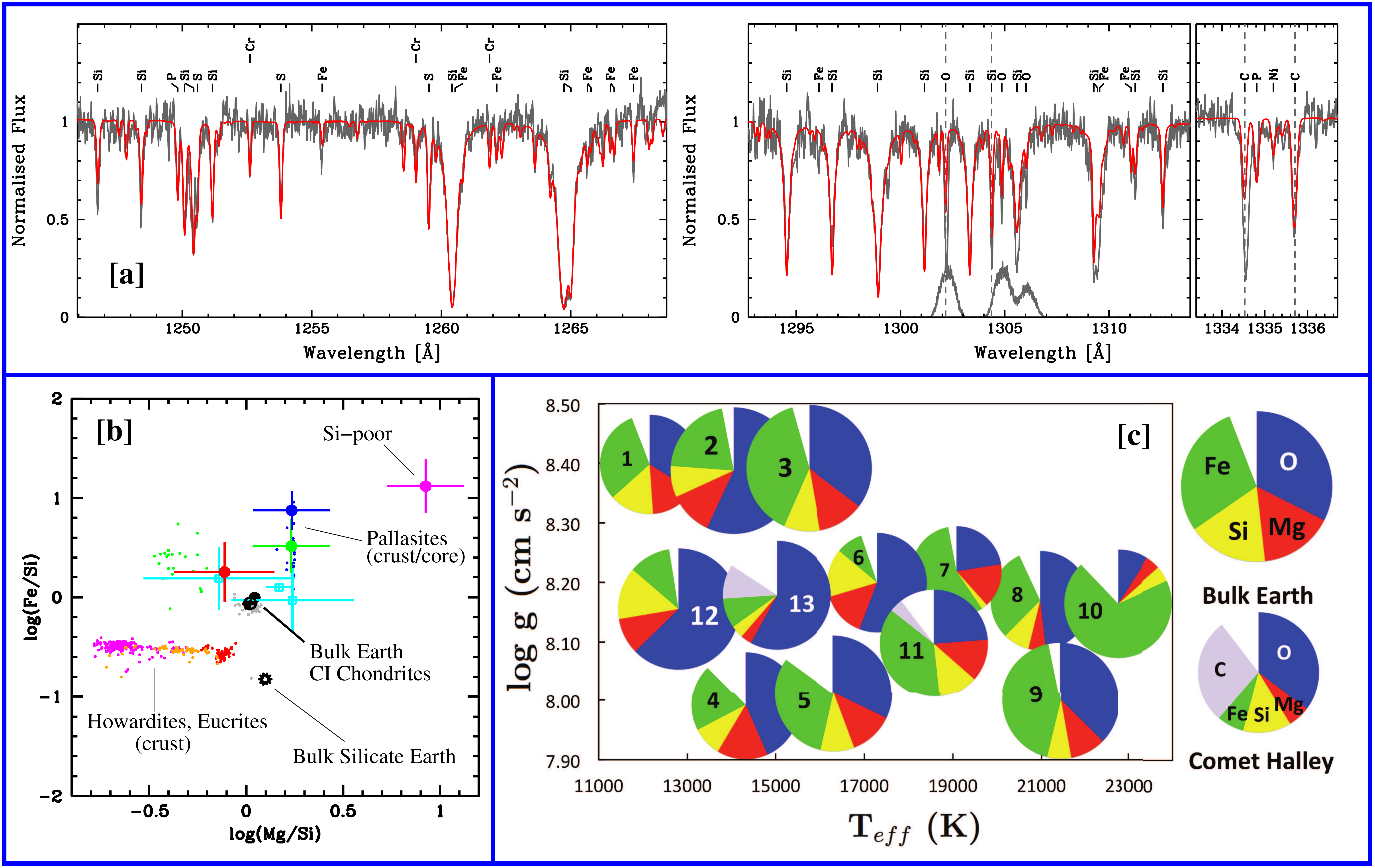}}
\vspace*{-4ex}
\textit{\caption{\label{f:wdplanets}
\textbf{[a]} Spectroscopy of white dwarfs accreting planetary debris \citep{gaensickeetal12-1} provides \textbf{[b]} accurate \textit{bulk compositions} of exo-planetesimals (small dots\,=\,solar-system meteorites, big dots\,=\,exo-planetesimals). Most published abundance studies are consistent with rocky compositions \citep{gaensickeetal12-1}, though there is evidence for water-rich asteroids \citep{farihietal13-1}, and comets \citep{xuetal17-1}. Detailed abundances have only been measured for a few systems  (\textbf{[c]}, \citealt{xuetal14-1}), limited by the small number of strongly metal-polluted white dwarfs accessible to ultraviolet spectroscopy with current facilities.}}
\end{figure*}

\medskip
\noindent
\textbf{Exo-planet bulk abundances across host star mass and age.}\\
With thousands of planets found, understanding their formation and evolution are now key research areas. The fundamental question \textit{``What are those other worlds made out of?''} is  difficult to answer from studies of planets orbiting main-sequence hosts, where radial velocities and transit light curves yield bulk densities. Further conclusions on internal structure and bulk composition are model-dependent \citep[e.g.][]{rogers+seager10-1, dornetal15-1}. \citet{zuckermanetal07-1} pioneered the spectroscopic analysis of white dwarfs accreting planetary debris to accurately measure the bulk composition of exo-planetary systems, analogous to solar-system meteorite studies, (Fig.\,\ref{f:wdplanets}). This method has been used to measure the abundances of rock-forming elements (Si, Fe, Mg, O), refractory lithophiles  (Ca, Al, Ti), siderophiles  (Cr, Mn, Ni), and volatiles, revealing a significant diversity (Fig.\,\ref{f:wdplanets}, bottom panels, \citealt{gaensickeetal12-1}) which includes evidence for differentiated planetesimals \citep{wilsonetal15-1, melisetal17-1}, water-rich exo-asteroids \citep{farihietal13-1, raddietal15-1} and one volatile-rich Kuiper belt-like body \citep{xuetal17-1}. This work provides important inputs into planet formation models \citep{carter-bondetal12-1, carteretal-2015}. 

After a phase of rapid progress, we are now limited by two factors: (1) The small number of known white dwarfs that are suitable for detailed abundance studies: while $\simeq$\,25\% of white dwarfs are weakly contaminated \citep{koesteretal14-1}, only $\simeq$1\% have accreted enough debris to enable the detection of multiple elements. (2) White dwarfs hotter than $\simeq15,000$\,K, and in particular those with opaque hydrogen atmospheres, require ultraviolet spectroscopy  (Fig\,\ref{f:wdplanets}, top panels) to carry out the abundance measurements, as the optical transitions rapidly weaken with increasing temperature. The first problem will be addressed thanks to \textit{Gaia}: \citet{gentile-fusilloetal19-1} used its second data release \citep{gaiaetal18-1} to compile the first homogeneously selected and practically complete sample of white dwarfs with a limiting magnitude of $G\simeq20$, comprising a staggering $\simeq$\,260,000 stars~--~a ten-fold increase over the previously number of known white dwarfs. However, the \textit{Gaia} data does not provide insight into the atmosphere compositions of these stars. Therefore optical spectroscopy of the entire sample is required to identify those few  thousand \textit{Gaia} white dwarfs that have accreted sufficiently large amounts of debris to warrant detailed abundance studies. High-resolution ultraviolet spectroscopy of these polluted white dwarfs obtained with a sensitive next-generation space telescope will then provided exquisite insight into the bulk compositions of exo-planetesimals. These abundance data will provide critically important inputs into planet formation models: (i) the volatile fraction traces the formation region of the planetesimals relative to the relevant condensation line, and objects originating from beyond the snow line \citep{xuetal17-1} provide insight into the bulk composition of the cores of gas giants, (ii) the Mg/Si ratio determines the composition of silicates with implications for planetary processes such as plate tectonics, and (iii) the relative abundances of Fe, siderophiles, and refractory lithophiles, provide insight into core and crust formation \citep{harrisonetal18-1}.

The detailed abundance studies of these systems will take the statistics of exo-planetesimal taxonomy to a level akin to that of solar system meteorite samples \citep{nittleretal04-1}. The progenitors of the \textit{Gaia} white dwarfs span masses of $M_\mathrm{ZAMS}\simeq1-8\,M_\odot$, and their total ages range from a few 100\,Myr to many Gyr, providing insight into the planet formation efficiency as a function of host mass and Galactic chemical evolution. Both metallicity and $\alpha$-element abundances of a star are expected to be a function of its age and formation location in the Galaxy \citep[e.g.][]{minchevetal13-1}. These differences are likely reflected in the composition of planetary systems that formed in different locations in the Galaxy and at different epochs, but subsequently migrated into the solar neighbourhood. The derived debris abundances of old planetary systems will be compared to the abundance trends seen in evolved planet hosts \citep{maldonado+villaver16-1}.

\begin{figure}
\includegraphics[width=\textwidth]{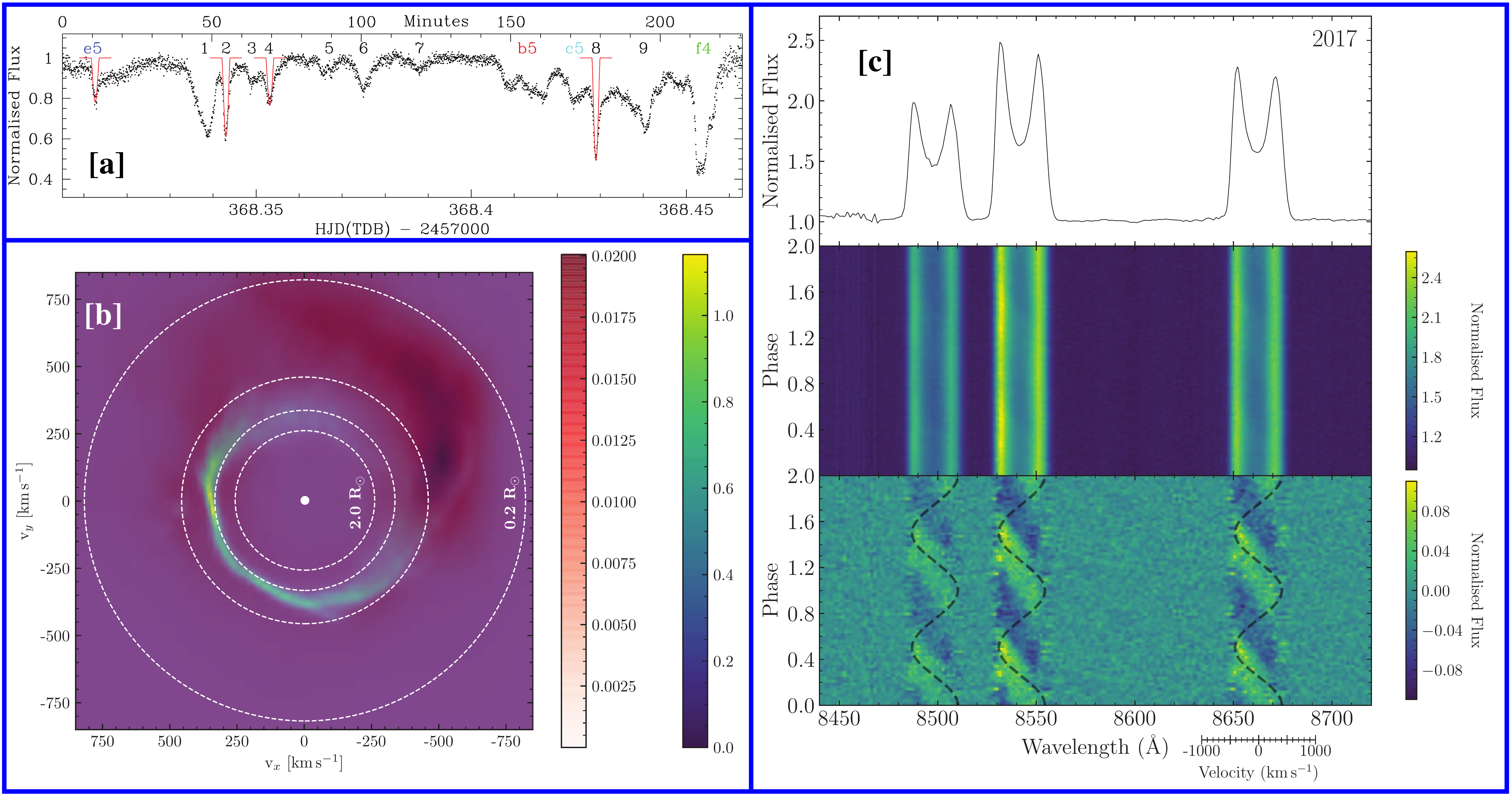}
\vspace*{-4ex}
\textit{\caption{Detections of planetesimals at white dwarfs. \textbf{[a]} High-speed (5s) ULTRASPEC photometry of WD1145+017  illustrating the complex structure of the transit events caused by planetary debris near the tidal disruption radius \citep{gaensickeetal16-1}. \textbf{[b]} The first image of a gaseous debris disk, showing an eccentric intensity pattern \citep{manseretal16-1}. \textbf{[c]} Fast GTC spectroscopy of this disk reveals a solid planetesimal with an orbital period of two hours \cite{manseretal19-1}, which is thought to produce the gas seen in the image on the left. \label{f:1145+1228}}}
\end{figure}

\medskip\noindent
\textbf{Identifying and characterising planetesimals around white dwarfs.}\\ 
The photometric detection of debris transits at WD\,1145+017 \citep{vanderburgetal15-1} and of the spectroscopic signature of a planetesimal orbiting in the gaseous disk at SDSS\,1228+1040 (\citealt{manseretal19-1}, Fig.\,\ref{f:1145+1228}) opened up the first opportunities to go beyond measuring the abundances of shredded debris, i.e. to characterise the physical properties of solid exo-planetary bodies. The intense follow-up of WD\,1145+017 \citep[e.g.][]{gaensickeetal16-1, rappaportetal16-1, redfieldetal17-1, izquierdoetal18-1} revealed a rapid evolution of the debris, providing real-time insight into the disintegration of a planetesimal deep within the gravitational potential of a white dwarf. Modelling the observations places robust constraints on the masses and internal structures of the  planetesimals at WD\,1145+017 and SDSS\,1228+1040 \citep{gurrietal17-1,verasetal17-2, manseretal19-1}. The physical make-up and the orbits of these objects also provide a more general insight into the architecture of remnant planetary systems. However, these are currently the only two exo-planetesimals known.

WD\,1145+017 was a serendipitous discovery, one of $\simeq$1,000 known white dwarfs observed by the \textit{Kepler/K2} mission. The strong debris pollution and its dust disk were found \textit{only after} the \textit{K2} detection of transits \citep{xuetal16-1}. Based on simple geometry, only a few per-cent of systems with close-in planetesimals will exhibit detectable transits~--~i.e. the discovery of transits at WD\,1145+017 is statistically consistent with \textit{all $\simeq$\,50 known white dwarfs with dust disks having close-in planetesimals}. While space-based photometry provides uninterrupted long time series and superb precision, neither of them are necessary to identify WD\,1145+017-like systems: the deep transits (reaching up to 50\%, see Fig.\,\ref{f:1145+1228}a) during the most active phase of WD\,1145+017 were easily detected with modest-sized telescopes from the ground \citep{garyetal17-1}, and even relatively sparse sampling will allow the identification of transits, if $\ga100$ epochs are obtained \citep{parsonsetal13-1}. Given that white dwarfs are sparsely distributed across the sky (even the \textit{Gaia} sample amounts only to $\simeq6$ per square degree) wide-area time domain surveys with cadences in the range of hours to days are ideal to identify additional white dwarfs with transiting debris. Detailed follow-up observations of these new systems require fast photometry with negligible overheads. Predictions based on a sample of one (WD\,1145+017) are naturally uncertain, but accounting for all evidence, several dozen such systems should be hiding among the \textit{Gaia} white dwarf sample.

Fast time-series spectroscopy of the gaseous debris disk at SDSS\,1228+1040 (Fig.\,\ref{f:1145+1228}b\,\&\,c) obtained with the 10.4\,m Gran Telescopio Canarias resulted in the detection of a planetesimal orbiting the white dwarf with a two-hour period. The significantly shorter period compared to WD\,1145+017 (4.5\,h) implies that this planetary body is solid with significant internal strength \citep{manseretal19-1}. Gaseous debris disks are extremely rare, only eight have been discovered so far, however, \textit{all but one} exhibit variability in the morphology of their \Ion{Ca}{ii} lines  \citep{manseretal16-2}. It is plausible that all gas disks are associated with close-in planetesimals, which we aim to confirm with deep time-series spectroscopy. \textit{This spectroscopic identification of planetesimals at white dwarfs is independent of their orbital inclination}. Seven of the eight known gas disks were identified from SDSS spectroscopy, which was incomplete in targeting white dwarfs and limited to $\simeq$\,1/3 of the sky. Scaling to the spectroscopic follow-up of the all-sky \textit{Gaia} white dwarf sample, at least $\simeq$\,30 gas disks are expected to be discovered. The search for planetesimals in these disks will require fast spectroscopic time-series on large-aperture telescopes equipped with medium-resolution optical spectrographs.  

Between photometric and spectroscopic detections, at least $\simeq50$ planetesimals in close orbits around white dwarfs will be found, and their physical properties will be determined following on from our previous work \citep{gurrietal17-1, verasetal17-2, manseretal19-1}. The range of masses and internal structures derived from this sample will provide the first major step into establishing the physical nature of the constituents of evolved planetary systems.

\medskip\noindent
\textbf{Infrared studies of debris disks variability and mineralogy}\\
Within the past two years, it has become clear that most dusty white dwarfs exhibit variable infrared emission due to as yet unconstrained dust production and removal processes \citep{swanetal19-1}.  This result relies heavily on the existence of warm {\em Spitzer} mission data, and less so on {\em WISE} and \textit{NEOWISE} due to sensitivity and source confusion \citep{xuetal18-1}.  Thus, it is clear that white dwarf planetary systems are active and novel science awaits the community but requires facilities that can sensitively measure micro-Jy fluxes at 3 and 4 microns with cadences of weeks to months to years.

A related and outstanding problem is the origin of the total planetary body mass that is observed via dust, gas, and metal pollution, which in some systems exceeds the mass of dwarf planets and solar systems moons \citep{juraetal09-2}. The consensus model involves the perturbation of minor bodies from an existing reservoir, but whether this reservoir is asteroid belt-like, Kuiper belt-like \citep{wyattetal14-1}, or generated in the post-main sequence \citep{verasetal14-2} is currently empirically unconstrained.  Thus the availability of a space mission sensitive in the  mid-infrared to far-infrared is critically important to definitively detect and characterize the belts of planetesimals that are implied by the dusty and gaseous debris disk systems.

Disks around polluted white dwarf stars have also been shown to exhibit strong solid-state emission features in their mid-infrared spectra \citep{reachetal05-1, reachetal09-1, juraetal09-1}. Characterising these features in high-definition will provide robust mineralogical information for polluting material before it is accreted by the host white dwarf star.

Combining far-ultraviolet and mid-infrared spectroscopic observations of dusty white dwarfs will provide bulk chemical compositions at an elemental level from (1) the stellar atmosphere and (2) dust stoichiometry from the disk.  This combined information will allow to comprehensively probe the formation and evolutionary history of the white dwarf's planetary system \citep[e.g.][]{bondetal10-1}. Such data will allow the identification of what specific chemical compounds parent  bodies were made of, thus enabling detailed physical modelling of the chemical, thermodynamic, and physical history of the accreted material.

\clearpage
\vspace*{-7ex}
\noindent
\textbf{Facility requirements for the next decade}\\
\noindent
\textbf{\textit{All-sky spectroscopic surveys.}}
Optical spectroscopy of the $260,000$ white dwarfs identified with \textit{Gaia} \citep{gentile-fusilloetal19-1} can only be achieved via piggy-backing on wide-area multi-object spectroscopic (MOS) surveys. With an average surface density of $\simeq6$ per square degree, observing all \textit{Gaia} white dwarfs will only take up a very small fraction of the fiber complement of modern MOS facilities. Coverage extending from $\simeq3800$\,\AA\ to $\simeq9000$\,\AA\ is essential to include the strongest optical metal transitions (Ca~H/K, \Line{Mg}{ii}{4481}\,\AA), and the \Line{Ca}{ii}{8600}\,\AA\ triplet which is seen in emission from gaseous debris disks. A spectral resolution of $\simeq4000$ is required to detect sharp metal lines, higher resolutions ($\simeq20,000$) will greatly increase the sensitivity to weak debris contamination, and metal pollution in hotter white dwarfs. 

\noindent
\textbf{\textit{High-resolution ultraviolet spectroscopy.}} The far-ultraviolet contains strong transitions of practically all elements relevant to diagnose the bulk composition and history of the accreted planetary bodies. \textit{HST}/COS has been the work horse for these studies so far, but is limited in sensitivity and spectral resolution. Model atmosphere analysis requires a signal-to-noise ratio of at least 30 which COS can achieve in a reasonable amount of orbits at $F\simeq10^{-14}\mathrm{erg\,cm^{-2}\,s^{-1}\,\AA^{-1}}$~--~only a few dozen of systems suitable for detailed abundance studies are sufficiently bright. Assuming a factor 30 increase in sensitivity compared to COS ($\times$15 for a 10\,m aperture, and $\times2$ from improved optics, and improved orbital visibility) will increase the available volume for detailed abundance studies by a factor $\simeq150$ compared to what can be reached with \textit{HST}, sufficient to include $>1000$ potential targets for high-quality ultraviolet spectroscopic follow-up. An increased spectral resolution, $\simeq50,000$ (compared to $10,000-20,000$ for COS) will greatly improve the modelling of blended lines, and is necessary to disentangle photospheric, circumstellar, and interstellar features. A wide  wavelength range, $\simeq900-3200$\,\AA, is desirable to include the higher Lyman lines which will (1) improve the accuracy of effective temperature and mass determinations, and in turn of the ages of the evolved planetary systems and (2) maximise the number of elements that can be used for the abundance measurements. 

\noindent
\textbf{\textit{Mid-infrared spectroscopy and imaging.}}
{\it James Webb Space Telescope} observations will provide the mid-infrared spectroscopic and imaging data required to address mineralogical and giant-planet companion science goals. Every white dwarf debris disk system currently known can be observed with high signal-to-noise (S/N$>$10) in $\lesssim$3~hours using the LRS module of MIRI, and several can even be observed profitably with its MRS module (e.g., Figure 9 of \citealt{dennihyetal16-1}). Imaging observations $-$ regardless of PSF-subtraction performance with the $Webb$ $-$ will enable planets down to sub-Saturn masses to be detected around white dwarfs.

\textit{JWST} will only partly address the detection of giant planets, high sensitivity in the mid- to far-infrared is required to identify the reservoirs that feed the debris-accreting white dwarfs, and the ability of flexible, multi-epoch observations are necessary to determine the nature and origin of the variability detected in many of the debris disks.

\noindent
\textbf{\textit{Wide-area time-domain surveys and fast photometry.}}
Identifying transiting debris at white dwarfs will require the ability to survey 10,000s of stars spread over the entire sky. Sparse (hours to days) sampling is sufficient, though stable long-term cadences will greatly facilitate the statistical analysis. Follow-up of new transiting systems requires fast (seconds, as white dwarfs are small and transits may last as short as a minute only) photometry, ideally in multiple bands simultaneously to probe for colour effects / extinction by the dusty debris.  

\noindent
\textbf{\textit{Large aperture telescope spectroscopy.}}
The spectroscopic detection of a planetesimal orbiting SDSS\,1228+1040 with a two-hour period demonstrates that solid bodies can achieve ultra-compact configurations and survive in these for considerable amounts of time. This study was carried out by obtaining fast ($\simeq$minutes) spectroscopy on the 10.4\,m GTC. While a few similarly bright gas disks are known, many of the new systems identified from the \textit{Gaia} follow-up will be $2-3$ magnitudes fainter, and probing for close-in planetesimals will require large aperture (30-40\,m) telescopes equipped with intermediate $\simeq5000$ resolution optical spectrographs. 

\clearpage

\bibliography{aamnem99,aabib}

\begin{thebibliography}{}
\makeatletter
\relax
\def\mn@urlcharsother{\let\do\@makeother \do\$\do\&\do\#\do\^\do\_\do\%\do\~}
\def\mn@doi{\begingroup\mn@urlcharsother \@ifnextchar [ {\mn@doi@}
  {\mn@doi@[]}}
\def\mn@doi@[#1]#2{\def\@tempa{#1}\ifx\@tempa\@empty \href
  {http://dx.doi.org/#2} {doi:#2}\else \href {http://dx.doi.org/#2} {#1}\fi
  \endgroup}
\def\mn@eprint#1#2{\mn@eprint@#1:#2::\@nil}
\def\mn@eprint@arXiv#1{\href {http://arxiv.org/abs/#1} {{\tt arXiv:#1}}}
\def\mn@eprint@dblp#1{\href {http://dblp.uni-trier.de/rec/bibtex/#1.xml}
  {dblp:#1}}
\def\mn@eprint@#1:#2:#3:#4\@nil{\def\@tempa {#1}\def\@tempb {#2}\def\@tempc
  {#3}\ifx \@tempc \@empty \let \@tempc \@tempb \let \@tempb \@tempa \fi \ifx
  \@tempb \@empty \def\@tempb {arXiv}\fi \@ifundefined
  {mn@eprint@\@tempb}{\@tempb:\@tempc}{\expandafter \expandafter \csname
  mn@eprint@\@tempb\endcsname \expandafter{\@tempc}}}

\bibitem[\protect\citeauthoryear{{Bond}, {O'Brien}  \& {Lauretta}}{{Bond}
  et~al.}{2010}]{bondetal10-1}
{Bond} J.~C.,  {O'Brien} D.~P.,   {Lauretta} D.~S.,  2010, \mn@doi [ApJ]
  {10.1088/0004-637X/715/2/1050}, \href {2010ApJ...715.1050B} {715, 1050}

\bibitem[\protect\citeauthoryear{{Carter-Bond}, {O'Brien}, {Delgado Mena},
  {Israelian}, {Santos}  \& {Gonz{\'a}lez Hern{\'a}ndez}}{{Carter-Bond}
  et~al.}{2012}]{carter-bondetal12-1}
{Carter-Bond} J.~C.,  {O'Brien} D.~P.,  {Delgado Mena} E.,  {Israelian} G.,
  {Santos} N.~C.,   {Gonz{\'a}lez Hern{\'a}ndez} J.~I.,  2012, \mn@doi [ApJ
  Lett.] {10.1088/2041-8205/747/1/L2}, \href {2012ApJ...747L...2C} {747, L2}

\bibitem[\protect\citeauthoryear{{Carter}, {Leinhardt}, {Elliott}, {Walter}  \&
  {Stewart}}{{Carter} et~al.}{2015}]{carteretal-2015}
{Carter} P.~J.,  {Leinhardt} Z.~M.,  {Elliott} T.,  {Walter} M.~J.,   {Stewart}
  S.~T.,  2015, \mn@doi [ApJ] {10.1088/0004-637X/813/1/72}, \href
  {2015ApJ...813...72C} {813, 72}

\bibitem[\protect\citeauthoryear{{Dennihy}, {Debes}, {Dunlap}, {Dufour},
  {Teske}  \& {Clemens}}{{Dennihy} et~al.}{2016}]{dennihyetal16-1}
{Dennihy} E.,  {Debes} J.~H.,  {Dunlap} B.~H.,  {Dufour} P.,  {Teske} J.~K.,
  {Clemens} J.~C.,  2016, \mn@doi [ApJ] {10.3847/0004-637X/831/1/31}, \href
  {2016ApJ...831...31D} {831, 31}

\bibitem[\protect\citeauthoryear{{Dorn}, {Khan}, {Heng}, {Connolly}, {Alibert},
  {Benz}  \& {Tackley}}{{Dorn} et~al.}{2015}]{dornetal15-1}
{Dorn} C.,  {Khan} A.,  {Heng} K.,  {Connolly} J.~A.~D.,  {Alibert} Y.,  {Benz}
  W.,   {Tackley} P.,  2015, \mn@doi [A\&A] {10.1051/0004-6361/201424915},
  \href {2015A&A...577A..83D} {577, A83}

\bibitem[\protect\citeauthoryear{{Farihi}}{{Farihi}}{2016}]{farihi16-1}
{Farihi} J.,  2016, \mn@doi [New Astronomy Reviews]
  {10.1016/j.newar.2016.03.001}, \href {2016NewAR..71....9F} {71, 9}

\bibitem[\protect\citeauthoryear{{Farihi}, {Jura}  \& {Zuckerman}}{{Farihi}
  et~al.}{2009}]{farihietal09-1}
{Farihi} J.,  {Jura} M.,   {Zuckerman} B.,  2009, \mn@doi [ApJ]
  {10.1088/0004-637X/694/2/805}, \href {2009ApJ...694..805F} {694, 805}

\bibitem[\protect\citeauthoryear{{Farihi}, {G{\"a}nsicke}  \&
  {Koester}}{{Farihi} et~al.}{2013}]{farihietal13-1}
{Farihi} J.,  {G{\"a}nsicke} B.~T.,   {Koester} D.,  2013, \mn@doi [MNRAS]
  {10.1093/mnras/stt432}, \href {2013MNRAS.432.1955F} {432, 1955}

\bibitem[\protect\citeauthoryear{{Gaia Collaboration} et~al.,}{{Gaia
  Collaboration} et~al.}{2018}]{gaiaetal18-1}
{Gaia Collaboration} et~al., 2018, \mn@doi [A\&A]
  {10.1051/0004-6361/201833051}, \href
  {http://adsabs.harvard.edu/abs/2018A%26A...616A...1G} {616, A1}

\bibitem[\protect\citeauthoryear{{G{\"a}nsicke}, {Marsh}, {Southworth}  \&
  {Rebassa-Mansergas}}{{G{\"a}nsicke} et~al.}{2006}]{gaensickeetal06-3}
{G{\"a}nsicke} B.~T.,  {Marsh} T.~R.,  {Southworth} J.,   {Rebassa-Mansergas}
  A.,  2006, \mn@doi [Science] {10.1126/science.1135033}, \href
  {2006Sci...314.1908G} {314, 1908}

\bibitem[\protect\citeauthoryear{{G{\"a}nsicke}, {Koester}, {Farihi}, {Girven},
  {Parsons}  \& {Breedt}}{{G{\"a}nsicke} et~al.}{2012}]{gaensickeetal12-1}
{G{\"a}nsicke} B.~T.,  {Koester} D.,  {Farihi} J.,  {Girven} J.,  {Parsons}
  S.~G.,   {Breedt} E.,  2012, \mn@doi [MNRAS]
  {10.1111/j.1365-2966.2012.21201.x}, \href {2012MNRAS.424..333G} {424, 333}

\bibitem[\protect\citeauthoryear{{G{\"a}nsicke} et~al.,}{{G{\"a}nsicke}
  et~al.}{2016}]{gaensickeetal16-1}
{G{\"a}nsicke} B.~T.,  et~al., 2016, \mn@doi [ApJ Lett.]
  {10.3847/2041-8205/818/1/L7}, \href {2016ApJ...818L...7G} {818, L7}

\bibitem[\protect\citeauthoryear{{Gary}, {Rappaport}, {Kaye}, {Alonso}  \&
  {Hambschs}}{{Gary} et~al.}{2017}]{garyetal17-1}
{Gary} B.~L.,  {Rappaport} S.,  {Kaye} T.~G.,  {Alonso} R.,   {Hambschs} F.-J.,
   2017, \mn@doi [MNRAS] {10.1093/mnras/stw2921}, \href {2017MNRAS.465.3267G}
  {465, 3267}

\bibitem[\protect\citeauthoryear{{Gentile Fusillo} et~al.,}{{Gentile Fusillo}
  et~al.}{2019}]{gentile-fusilloetal19-1}
{Gentile Fusillo} N.~P.,  et~al., 2019, \mn@doi [MNRAS]
  {10.1093/mnras/sty3016}, \href {2019MNRAS.482.4570G} {482, 4570}

\bibitem[\protect\citeauthoryear{{Gurri}, {Veras}  \& {G{\"a}nsicke}}{{Gurri}
  et~al.}{2017}]{gurrietal17-1}
{Gurri} P.,  {Veras} D.,   {G{\"a}nsicke} B.~T.,  2017, \mn@doi [MNRAS]
  {10.1093/mnras/stw2293}, \href {2017MNRAS.464..321G} {464, 321}

\bibitem[\protect\citeauthoryear{{Harrison}, {Bonsor}  \&
  {Madhusudhan}}{{Harrison} et~al.}{2018}]{harrisonetal18-1}
{Harrison} J.~H.~D.,  {Bonsor} A.,   {Madhusudhan} N.,  2018, \mn@doi [MNRAS]
  {10.1093/mnras/sty1700}, \href {2018MNRAS.479.3814H} {479, 3814}

\bibitem[\protect\citeauthoryear{{Izquierdo} et~al.,}{{Izquierdo}
  et~al.}{2018}]{izquierdoetal18-1}
{Izquierdo} P.,  et~al., 2018, \mn@doi [MNRAS] {10.1093/mnras/sty2315}, \href
  {2018MNRAS.481..703I} {481, 703}

\bibitem[\protect\citeauthoryear{{Jura}}{{Jura}}{2003}]{jura03-1}
{Jura} M.,  2003, \mn@doi [ApJ Lett.] {10.1086/374036}, \href
  {2003ApJ...584L..91J} {584, L91}

\bibitem[\protect\citeauthoryear{{Jura}, {Farihi}  \& {Zuckerman}}{{Jura}
  et~al.}{2009a}]{juraetal09-1}
{Jura} M.,  {Farihi} J.,   {Zuckerman} B.,  2009a, \mn@doi [AJ]
  {10.1088/0004-6256/137/2/3191}, \href {2009AJ....137.3191J} {137, 3191}

\bibitem[\protect\citeauthoryear{{Jura}, {Muno}, {Farihi}  \&
  {Zuckerman}}{{Jura} et~al.}{2009b}]{juraetal09-2}
{Jura} M.,  {Muno} M.~P.,  {Farihi} J.,   {Zuckerman} B.,  2009b, \mn@doi [ApJ]
  {10.1088/0004-637X/699/2/1473}, \href {2009ApJ...699.1473J} {699, 1473}

\bibitem[\protect\citeauthoryear{{Koester}, {G{\"a}nsicke}  \&
  {Farihi}}{{Koester} et~al.}{2014}]{koesteretal14-1}
{Koester} D.,  {G{\"a}nsicke} B.~T.,   {Farihi} J.,  2014, \mn@doi [A\&A]
  {10.1051/0004-6361/201423691}, \href {2014A&A...566A..34K} {566, A34}

\bibitem[\protect\citeauthoryear{{Maldonado} \& {Villaver}}{{Maldonado} \&
  {Villaver}}{2016}]{maldonado+villaver16-1}
{Maldonado} J.,  {Villaver} E.,  2016, \mn@doi [A\&A]
  {10.1051/0004-6361/201527883}, \href {2016A&A...588A..98M} {588, A98}

\bibitem[\protect\citeauthoryear{{Manser} et~al.,}{{Manser}
  et~al.}{2016a}]{manseretal16-1}
{Manser} C.~J.,  et~al., 2016a, \mn@doi [MNRAS] {10.1093/mnras/stv2603}, \href
  {2016MNRAS.455.4467M} {455, 4467}

\bibitem[\protect\citeauthoryear{{Manser}, {G{\"a}nsicke}, {Koester}, {Marsh}
  \& {Southworth}}{{Manser} et~al.}{2016b}]{manseretal16-2}
{Manser} C.~J.,  {G{\"a}nsicke} B.~T.,  {Koester} D.,  {Marsh} T.~R.,
  {Southworth} J.,  2016b, \mn@doi [MNRAS] {10.1093/mnras/stw1760}, \href
  {2016MNRAS.462.1461M} {462, 1461}

\bibitem[\protect\citeauthoryear{{Manser} et~al.,}{{Manser}
  et~al.}{2019}]{manseretal19-1}
{Manser} C.~J.,  et~al., 2019, Science, 364, 66

\bibitem[\protect\citeauthoryear{{Melis} \& {Dufour}}{{Melis} \&
  {Dufour}}{2017}]{melisetal17-1}
{Melis} C.,  {Dufour} P.,  2017, \mn@doi [ApJ] {10.3847/1538-4357/834/1/1},
  \href {2017ApJ...834....1M} {834, 1}

\bibitem[\protect\citeauthoryear{{Minchev}, {Chiappini}  \& {Martig}}{{Minchev}
  et~al.}{2013}]{minchevetal13-1}
{Minchev} I.,  {Chiappini} C.,   {Martig} M.,  2013, \mn@doi [A\&A]
  {10.1051/0004-6361/201220189}, \href {2013A&A...558A...9M} {558, A9}

\bibitem[\protect\citeauthoryear{{Nittler}, {McCoy}, {Clark}, {Murphy},
  {Trombka}  \& {Jarosewich}}{{Nittler} et~al.}{2004}]{nittleretal04-1}
{Nittler} L.~R.,  {McCoy} T.~J.,  {Clark} P.~E.,  {Murphy} M.~E.,  {Trombka}
  J.~I.,   {Jarosewich} E.,  2004, Antarctic Meteorite Research, \href
  {2004AMR....17..231N} {17, 231}

\bibitem[\protect\citeauthoryear{{Parsons}, {Marsh}, {G{\"a}nsicke},
  {Schreiber}, {Bours}, {Dhillon}  \& {Littlefair}}{{Parsons}
  et~al.}{2013}]{parsonsetal13-1}
{Parsons} S.~G.,  {Marsh} T.~R.,  {G{\"a}nsicke} B.~T.,  {Schreiber} M.~R.,
  {Bours} M.~C.~P.,  {Dhillon} V.~S.,   {Littlefair} S.~P.,  2013, \mn@doi
  [MNRAS] {10.1093/mnras/stt1588}, \href {2013MNRAS.436..241P} {436, 241}

\bibitem[\protect\citeauthoryear{{Payne}, {Veras}, {G{\"a}nsicke}  \&
  {Holman}}{{Payne} et~al.}{2017}]{payneetal17-1}
{Payne} M.~J.,  {Veras} D.,  {G{\"a}nsicke} B.~T.,   {Holman} M.~J.,  2017,
  \mn@doi [MNRAS] {10.1093/mnras/stw2585}, \href {2017MNRAS.464.2557P} {464,
  2557}

\bibitem[\protect\citeauthoryear{{Raddi}, {G{\"a}nsicke}, {Koester}, {Farihi},
  {Hermes}, {Scaringi}, {Breedt}  \& {Girven}}{{Raddi}
  et~al.}{2015}]{raddietal15-1}
{Raddi} R.,  {G{\"a}nsicke} B.~T.,  {Koester} D.,  {Farihi} J.,  {Hermes}
  J.~J.,  {Scaringi} S.,  {Breedt} E.,   {Girven} J.,  2015, \mn@doi [MNRAS]
  {10.1093/mnras/stv701}, \href {2015MNRAS.450.2083R} {450, 2083}

\bibitem[\protect\citeauthoryear{{Rappaport}, {Gary}, {Kaye}, {Vanderburg},
  {Croll}, {Benni}  \& {Foote}}{{Rappaport} et~al.}{2016}]{rappaportetal16-1}
{Rappaport} S.,  {Gary} B.~L.,  {Kaye} T.,  {Vanderburg} A.,  {Croll} B.,
  {Benni} P.,   {Foote} J.,  2016, \mn@doi [MNRAS] {10.1093/mnras/stw612},
  \href {2016MNRAS.458.3904R} {458, 3904}

\bibitem[\protect\citeauthoryear{{Reach}, {Kuchner}, {von Hippel}, {Burrows},
  {Mullally}, {Kilic}  \& {Winget}}{{Reach} et~al.}{2005}]{reachetal05-1}
{Reach} W.~T.,  {Kuchner} M.~J.,  {von Hippel} T.,  {Burrows} A.,  {Mullally}
  F.,  {Kilic} M.,   {Winget} D.~E.,  2005, \mn@doi [ApJ Lett.]
  {10.1086/499561}, \href {2005ApJ...635L.161R} {635, L161}

\bibitem[\protect\citeauthoryear{{Reach}, {Lisse}, {von Hippel}  \&
  {Mullally}}{{Reach} et~al.}{2009}]{reachetal09-1}
{Reach} W.~T.,  {Lisse} C.,  {von Hippel} T.,   {Mullally} F.,  2009, \mn@doi
  [ApJ] {10.1088/0004-637X/693/1/697}, \href {2009ApJ...693..697R} {693, 697}

\bibitem[\protect\citeauthoryear{{Redfield}, {Farihi}, {Cauley}, {Parsons},
  {G{\"a}nsicke}  \& {Duvvuri}}{{Redfield} et~al.}{2017}]{redfieldetal17-1}
{Redfield} S.,  {Farihi} J.,  {Cauley} P.~W.,  {Parsons} S.~G.,  {G{\"a}nsicke}
  B.~T.,   {Duvvuri} G.~M.,  2017, \mn@doi [ApJ] {10.3847/1538-4357/aa68a0},
  \href {2017ApJ...839...42R} {839, 42}

\bibitem[\protect\citeauthoryear{{Rogers} \& {Seager}}{{Rogers} \&
  {Seager}}{2010}]{rogers+seager10-1}
{Rogers} L.~A.,  {Seager} S.,  2010, \mn@doi [ApJ]
  {10.1088/0004-637X/712/2/974}, \href {2010ApJ...712..974R} {712, 974}

\bibitem[\protect\citeauthoryear{{Schr{\"o}der} \& {Connon
  Smith}}{{Schr{\"o}der} \& {Connon Smith}}{2008}]{schroeder+connonsmith08-1}
{Schr{\"o}der} K.,  {Connon Smith} R.,  2008, \mn@doi [MNRAS]
  {10.1111/j.1365-2966.2008.13022.x}, \href {2008MNRAS.386..155S} {386, 155}

\bibitem[\protect\citeauthoryear{{Swan}, {Farihi}  \& {Wilson}}{{Swan}
  et~al.}{2019}]{swanetal19-1}
{Swan} A.,  {Farihi} J.,   {Wilson} T.~G.,  2019, \mn@doi [MNRAS]
  {10.1093/mnrasl/slz014}, \href {2019MNRAS.484L.109S} {484, L109}

\bibitem[\protect\citeauthoryear{{Vanderburg} et~al.,}{{Vanderburg}
  et~al.}{2015}]{vanderburgetal15-1}
{Vanderburg} A.,  et~al., 2015, \mn@doi [Nat] {10.1038/nature15527}, \href
  {2015Natur.526..546V} {526, 546}

\bibitem[\protect\citeauthoryear{{Veras}}{{Veras}}{2016}]{veras16-1}
{Veras} D.,  2016, Royal Society Open Science

\bibitem[\protect\citeauthoryear{{Veras} \& {G{\"a}nsicke}}{{Veras} \&
  {G{\"a}nsicke}}{2015}]{veras+gaensicke15-1}
{Veras} D.,  {G{\"a}nsicke} B.~T.,  2015, \mn@doi [MNRAS]
  {10.1093/mnras/stu2475}, \href {2015MNRAS.447.1049V} {447, 1049}

\bibitem[\protect\citeauthoryear{{Veras}, {Jacobson}  \&
  {G{\"a}nsicke}}{{Veras} et~al.}{2014}]{verasetal14-2}
{Veras} D.,  {Jacobson} S.~A.,   {G{\"a}nsicke} B.~T.,  2014, \mn@doi [MNRAS]
  {10.1093/mnras/stu1926}, \href {2014MNRAS.445.2794V} {445, 2794}

\bibitem[\protect\citeauthoryear{{Veras}, {Carter}, {Leinhardt}  \&
  {G{\"a}nsicke}}{{Veras} et~al.}{2017}]{verasetal17-2}
{Veras} D.,  {Carter} P.~J.,  {Leinhardt} Z.~M.,   {G{\"a}nsicke} B.~T.,  2017,
  \mn@doi [MNRAS] {10.1093/mnras/stw2748}, \href {2017MNRAS.465.1008V} {465,
  1008}

\bibitem[\protect\citeauthoryear{{Wilson}, {G{\"a}nsicke}, {Koester}, {Toloza},
  {Pala}, {Breedt}  \& {Parsons}}{{Wilson} et~al.}{2015}]{wilsonetal15-1}
{Wilson} D.~J.,  {G{\"a}nsicke} B.~T.,  {Koester} D.,  {Toloza} O.,  {Pala}
  A.~F.,  {Breedt} E.,   {Parsons} S.~G.,  2015, \mn@doi [MNRAS]
  {10.1093/mnras/stv1201}, \href {2015MNRAS.451.3237W} {451, 3237}

\bibitem[\protect\citeauthoryear{{Wyatt}, {Farihi}, {Pringle}  \&
  {Bonsor}}{{Wyatt} et~al.}{2014}]{wyattetal14-1}
{Wyatt} M.~C.,  {Farihi} J.,  {Pringle} J.~E.,   {Bonsor} A.,  2014, \mn@doi
  [MNRAS] {10.1093/mnras/stu183}, \href {2014MNRAS.439.3371W} {439, 3371}

\bibitem[\protect\citeauthoryear{{Xu}, {Jura}, {Koester}, {Klein}  \&
  {Zuckerman}}{{Xu} et~al.}{2014}]{xuetal14-1}
{Xu} S.,  {Jura} M.,  {Koester} D.,  {Klein} B.,   {Zuckerman} B.,  2014,
  \mn@doi [ApJ] {10.1088/0004-637X/783/2/79}, \href {2014ApJ...783...79X} {783,
  79}

\bibitem[\protect\citeauthoryear{{Xu}, {Jura}, {Dufour}  \& {Zuckerman}}{{Xu}
  et~al.}{2016}]{xuetal16-1}
{Xu} S.,  {Jura} M.,  {Dufour} P.,   {Zuckerman} B.,  2016, \mn@doi [ApJ Lett.]
  {10.3847/2041-8205/816/2/L22}, \href {2016ApJ...816L..22X} {816, L22}

\bibitem[\protect\citeauthoryear{{Xu}, {Zuckerman}, {Dufour}, {Young}, {Klein}
  \& {Jura}}{{Xu} et~al.}{2017}]{xuetal17-1}
{Xu} S.,  {Zuckerman} B.,  {Dufour} P.,  {Young} E.~D.,  {Klein} B.,   {Jura}
  M.,  2017, \mn@doi [ApJ Lett.] {10.3847/2041-8213/836/1/L7}, \href
  {2017ApJ...836L...7X} {836, L7}

\bibitem[\protect\citeauthoryear{{Xu} et~al.,}{{Xu} et~al.}{2018}]{xuetal18-1}
{Xu} S.,  et~al., 2018, \mn@doi [MNRAS] {10.1093/mnras/stx3023}, \href
  {2018MNRAS.474.4795X} {474, 4795}

\bibitem[\protect\citeauthoryear{{Zuckerman}, {Koester}, {Melis}, {Hansen}  \&
  {Jura}}{{Zuckerman} et~al.}{2007}]{zuckermanetal07-1}
{Zuckerman} B.,  {Koester} D.,  {Melis} C.,  {Hansen} B.~M.,   {Jura} M.,
  2007, \mn@doi [ApJ] {10.1086/522223}, \href {2007ApJ...671..872Z} {671, 872}

\bibitem[\protect\citeauthoryear{{Zuckerman}, {Melis}, {Klein}, {Koester}  \&
  {Jura}}{{Zuckerman} et~al.}{2010}]{zuckermanetal10-1}
{Zuckerman} B.,  {Melis} C.,  {Klein} B.,  {Koester} D.,   {Jura} M.,  2010,
  \mn@doi [ApJ] {10.1088/0004-637X/722/1/725}, \href {2010ApJ...722..725Z}
  {722, 725}

\makeatother
\end{thebibliography}
\bibliographystyle{mnras}

\end{document}